\documentclass[prl,aps,twocolumn,showpacs, floatfix]{revtex4}
\usepackage{amssymb}
\usepackage{hyperref}
\usepackage{graphicx}
\usepackage{dcolumn}
\usepackage{bm,amsmath,verbatim}
\usepackage{mathrsfs}

\def\be{\begin{equation}}
\def\ee{\end{equation}}
\def\bea{\begin{eqnarray}}
\def\eea{\end{eqnarray}}
\def\bse{\begin{subequations}}
\def\ese{\end{subequations}}

\def\be{\begin{eqnarray}}
\def\ee{\end{eqnarray}}

\begin{document}

\title{Dark Solitons with Majorana Fermions in Spin-Orbit-Coupled Fermi Gases%
}
\author{Yong Xu$^{1}$}
\thanks{These authors contributed equally to the work}
\author{Li Mao$^{2,1}$}
\thanks{These authors contributed equally to the work}
\author{Biao Wu$^{3,4}$}
\author{Chuanwei Zhang$^{1}$}
\thanks{Corresponding Author, Email: chuanwei.zhang@utdallas.edu}
\affiliation{$^{1}$Department of Physics, The University of Texas at Dallas, Richardson,
Texas 75080, USA}
\affiliation{$^{2}$School of Physics and Technology, Wuhan University, Wuhan 430072, China}
\affiliation{$^{3}$International Center for Quantum Materials, School of Physics, Peking University, Beijing
100871, China}
\affiliation{$^{4}$Collaborative Innovation Center of Quantum Matter, Beijing 100871, China}

\begin{abstract}
We show that a single dark soliton can exist in a spin-orbit-coupled Fermi
gas with a high spin imbalance, where spin-orbit coupling favors uniform
superfluids over non-uniform Fulde-Ferrell-Larkin-Ovchinnikov states,
leading to dark soliton excitations in highly imbalanced gases. Above a
critical spin imbalance, two topological Majorana fermions (MFs) without
interactions can coexist inside a dark soliton, paving a way for
manipulating MFs through controlling solitons. At the topological transition
point, the atom density contrast across the soliton suddenly vanishes,
suggesting a signature for identifying topological solitons.
\end{abstract}

\maketitle

Solitons, topological defects arising from the interplay between dispersion
and nonlinearity of underlying systems, are significant for many different
physical branches~\cite{DrazinBook,DickeyBook,KivsharBook,KevrekidisBook}.
The realization of cold atomic superfluids provides a clean and controllable
platform for exploring soliton physics. In cold atomic gases, dark solitons
represent quantum excitations of a superfluid with the superfluid order
parameter vanishing at the soliton center in conjunction with a phase jump
across the soliton. Dark solitons have been extensively investigated in cold
atoms \cite{Carr2000JPB,Jackson1998PRA,
Dum1998PRL,Muryshev1999PRA,Fedichev1999PRA,DSOCEPL2013,Burger1999PRL,Denschlag2000Science,Anderson2001PRL}%
. In particular, dark solitons have recently been experimentally observed in
strongly interacting spin balanced Fermi gases~\cite{Martin2013Nature},
where the Cooper pairing wave-function has a phase jump across the soliton~%
\cite{Dziarmaga2005Laser,Stringari2007PRA,Stringari2012PRL,Liao2011PRA,
Spuntarelli2011NJP,Scott2012NJP,Cetoli2013PRA}. However, dark solitons in
the presence of a large spin imbalance have not been well explored.

With a large spin imbalance, the ground state of the superfluid is
theoretically predicted to be the spatially non-uniform
Fulde-Ferrell-Larkin-Ovchinnikov (FFLO) phase~\cite{FFLOlit} with finite
momentum pairing in 1D and quasi-1D~\cite%
{Yoshida2007PRA,HuHui2007PRA,Parish2007PRL,Roman2011PRA,Kuei2012PRA,Xiwen2013RMP}%
, which is partially verified by the experiment~\cite{Hulet2010Nature}. This
spatially non-uniform phase does not support dark soliton excitations that
usually occur in BCS-type uniform superfluids. On the other hand, uniform
superfluids can exist in large spin imbalanced Fermi gases~\cite{QuArxiv} in
the presence of spin-orbit (SO) coupling. Since SO coupling for cold atoms
has been experimentally generated recently for both bosons and fermions \cite%
{Lin2011Nature,Jing2012PRL,Zwierlen2012PRL,PanJian2012PRL,
Qu2013PRA,Spilman2013PRL}, a natural question is whether such SO coupled
superfluids with large spin imbalances can also support dark solitons.

More interestingly, it is well known that defects (vortices, edges, etc.) in
SO coupled fermionic superfluids with large spin imbalances can accommodate
Majorana fermions (MFs) \cite{Zhang2008PRL,Sato2009PRL,ShiLiang2011PRL,
LJiang2011PRL,XJLiu2012PRA,Melo2012PRL}, topological excitations that
satisfy exotic non-Abelian exchange statistics~\cite{Sarma2008RMP}. Recently
MFs have attracted tremendous attention in various physical systems \cite%
{MFSystem} because of their fundamental importance as well as potential
applications in fault-tolerant quantum computation~\cite{Kitaev2003AP}. In
this context, SO coupled fermionic superfluids have their intrinsic
advantages for MFs because of their disorder free~\cite{Liu2012PRL} and
highly controllable characteristics. Therefore another important question is
whether topological Majorana excitations can exist inside dark solitons if
such topological defects do exist.

In this Letter, we address these two important questions by studying dark
solitons in degenerate Fermi gases (DFGs) trapped in 1D harmonic potentials
with the experimentally already realized SO coupling and spin imbalances.
Here the spin imbalance is equivalent to a Zeeman field. In the absence of
SO coupling, the FFLO state~\cite{FFLOlit} with an oscillating order
parameter amplitude is the ground state \cite{HuHui2007PRA} with a large
Zeeman field, which cannot support dark solitons. With SO coupling, we find

(i) SO coupling suppresses the FFLO state, leading to uniform BCS
superfluids that support dark solitons. The parameter region for dark
solitons and their spatial properties are obtained.

(ii) For substantially large spin imbalances, we find remarkably that two
MFs can coexist inside a dark soliton without any interaction, beyond the
general expectation that two MFs with overlapping wave-functions interact,
leading to energy splitting that destroys MFs. Such solitons are topological
solitons to be distinguished from solitons without MFs.

(iii) The experimental signature of MFs inside the dark soliton in the local
density of states (LDOS) is characterized, which show an isolated zero
energy peak at the center of the soliton. Moreover, the density contrast
across the soliton suddenly decreases to zero at the topological transition
point, which may be used to experimentally detect topological solitons.

\emph{System and Hamiltonian}: Consider a SO coupled DFG confined in a 1D
harmonic trap with the transversal confinement provided by a tightly focused
optical dipole trap. The many-body Hamiltonian of the system can be written
as
\begin{eqnarray}
H &=&\int d{x}\hat{\Psi}^{\dagger }({x})H_{s}\hat{\Psi}({x}) \\
&&-g\int d{x}\hat{\Psi}_{\uparrow }^{\dagger }({x})\hat{\Psi}_{\downarrow
}^{\dagger }({x})\hat{\Psi}_{\downarrow }({x})\hat{\Psi}_{\uparrow }({x}),
\notag
\end{eqnarray}%
where the single particle grand-canonical Hamiltonian $H_{s}=-\hbar
^{2}\partial _{x}^{2}/2m-\mu +V({x})+H_{\text{SOC}}+H_{z}$, the harmonic
trapping potential $V({x})=m\omega ^{2}x^{2}/2$, $\mu $ is the chemical
potential, $g$ is the attractive s-wave scattering interaction strength
between atoms that can be tuned through Feshbach resonances, $m$ is the atom
mass, and $\omega $ is the trapping frequency. $\hat{\Psi}({x})=[\hat{\Psi}%
_{\uparrow }({x}),\hat{\Psi}_{\downarrow }({x})]^{T}$ with the atom creation
(annihilation) operator $\hat{\Psi}_{\nu }^{\dagger }({x})$ ($\hat{\Psi}%
_{\nu }({x})$) at spin $\nu $ and position $x$. We consider the equal Rashba
and Dresselhaus SO coupling $H_{SOC}=-i\hbar \alpha \partial _{x}\sigma _{y}$%
, where $\sigma _{i}$ are Pauli matrices. The Zeeman field $%
H_{z}=V_{z}\sigma _{z}$ generates the spin imbalance. This type of SO
coupling and Zeeman field have been realized experimentally~\cite%
{Lin2011Nature,Jing2012PRL,Zwierlen2012PRL,PanJian2012PRL,
Qu2013PRA,Spilman2013PRL} for cold atom Fermi gases using two
counter-propagating Raman lasers that couple two atomic hyperfine ground
states (i.e., the spin).

Within the standard mean-field approximation, the fermionic superfluids can
be described by the Bogoliubov-de Gennes (BdG) equation
\begin{equation}
\left(
\begin{array}{cc}
H_{s} & \Delta ({x}) \\
\Delta ({x})^{\ast } & -\sigma _{y}H_{s}^{\ast }\sigma _{y}%
\end{array}%
\right) \left(
\begin{array}{c}
u_{n} \\
v_{n}%
\end{array}%
\right) =E_{n}\left(
\begin{array}{c}
u_{n} \\
v_{n}%
\end{array}%
\right) ,  \label{BdGE}
\end{equation}%
where $u_{n}=[u_{n\uparrow }(x),u_{n\downarrow }(x)]^{T}$, $%
v_{n}=[v_{n\downarrow }(x),-v_{n\uparrow }(x)]$ are the Nambu spinor
wave-functions for the quasi-particle excitation energy $E_{n}$, the order
parameter $\Delta ({x})=-g\langle \hat{\Psi}_{\downarrow }({x})\hat{\Psi}%
_{\uparrow }({x})\rangle =-(g/2)\sum_{|E_{n}|<E_{c}}\left[ u_{n\uparrow
}v_{n\downarrow }^{\ast }f(E_{n})+u_{n\downarrow }v_{n\uparrow }^{\ast
}f(-E_{n})\right] $, and the atom density $\rho _{\sigma }({x}%
)=(1/2)\sum_{|E_{n}|<E_{c}}\left[ |u_{n\sigma }|^{2}f(E_{n})+|v_{n\sigma
}|^{2}f(-E_{n})\right] $ with the energy cut-off $E_{c}$. $%
f(E)=1/(1+e^{E/k_{B}T})$ is the quasi-particle Fermi-Dirac distribution at
the temperature $T$. With the constraints of a fixed total number of atoms $%
N=\int d{x}\left[ \rho _{\uparrow }({x})+\rho _{\downarrow }({x})\right] $
and the definition of the order parameter, Eq.~(\ref{BdGE}) can be solved
self-consistently. To obtain a stationary soliton excitation in the
superfluid, we choose $\Delta \tanh (x/\xi)$ with coherent length $\xi
=\hbar v_{F}/\Delta $ and Fermi velocity $v_{F}=\hbar K_{F}/m$ as the
initial order parameter, and then solve the BdG equations self-consistently
until the order parameter and density converge.

To solve the BdG equation, we expand $u_{n}$ and $v_{n}$ on the basis states
of the harmonic oscillator to convert the equation to a diagonalization
problem of a secular matrix. We consider $N=100$ atoms with 300 harmonic
oscillator states for the wave-function expansion. The energy cut-off $%
E_{c}=240\hbar \omega $, which is large enough to ensure the accuracy of the
calculation~\cite{HuHui2007PRA}. We choose the single particle Fermi energy $%
E_{F}=N\hbar \omega /2$ (neglecting zero point energy) in the absence of the
SO coupling and Zeeman field and the harmonic oscillator length $x_{s}=\sqrt{%
\hbar /m\omega }$ as the units of energy and length. For 1D Fermi gases, the
interaction parameter $g=-2\hbar ^{2}/ma_{1D}$ with an effective 1D
scattering length $a_{1D}$~\cite{Hulet2010Nature}. A dimensionless parameter
$\Gamma =-mg/n(0)\hbar ^{2}=\pi x_{s}/\sqrt{N}a_{1D}$, which is proportional
to the ratio between the interaction and kinetic energy at the center, can
be used to characterize the interaction strength~\cite{HuHui2007PRA}. Here $%
n(0)$ is the density at the trap center in Thomas-Fermi approximation. In
experiments~\cite{Hulet2010Nature}, this value can be as large as $5.2$. We
choose $\Gamma =\pi $ in most of our calculations.

\begin{figure}[t]
\includegraphics[width=3.4in]{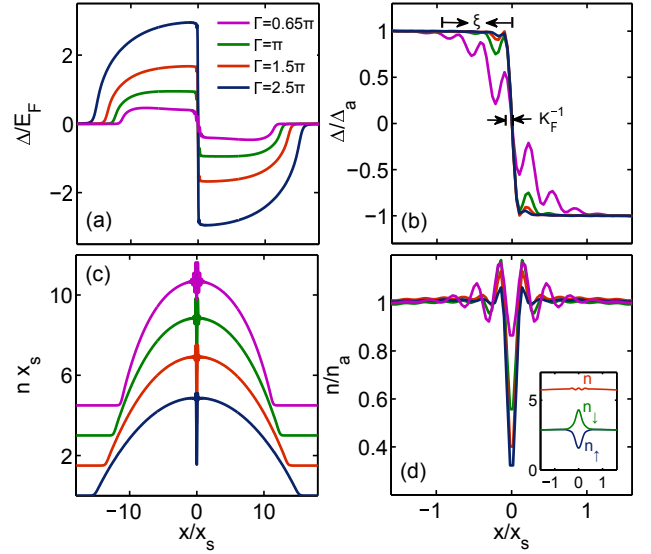}
\caption{(Color online) Profiles of the order parameter $\Delta $ ((a) and
(b)) and the total density $n$ ((c) and (d)) with increasing atom
interactions. In (b) and (d), the details of dark solitons in a small region
are plotted, where $\Delta $ and $n$ are respectively scaled by $\Delta _{a}$
and $n_{a}$, which are the asymptotic value to the origin point without the
soliton. In (c), lines corresponding to $\Gamma =1.5\protect\pi ,\protect\pi %
,0.65\protect\pi $ are moved up $1.5x_{s}, 3.0x_s, 4.5x_s$ with respect to
the line for $\Gamma=2.5\protect\pi$. In the inset of (d), the density
profiles ($n_\protect\sigma$ for spin $\protect\sigma$ and $%
n=n_{\uparrow}+n_{\downarrow}$) for a gas with $\Gamma=\protect\pi$ without
SO coupling are plotted. Here $\protect\alpha k_{F}=E_{F}$ and $%
V_{z}=0.16E_{F}$. }
\label{Delta_V8}
\end{figure}

\emph{Dark solitons in spin imbalanced DFGs}: In Fig.~\ref{Delta_V8}, we
plot the order parameter and density profiles for an imbalanced SO coupled
Fermi gas with different interaction strengths. With increasing
interactions, both $\Delta $ and the depth of the soliton increase while $%
\mu $ (not shown here) decreases, signalling the crossover from BCS
superfluids to BEC molecule bound states. Without SO coupling, the existence
of a dark soliton leads to the depletion (enhancement) of spin $\uparrow $ ($%
\downarrow $) component even with small Zeeman fields, thus the total atom
density only has a small depletion at the soliton center as shown in the
inset of Fig.~\ref{Delta_V8}(d). This is in sharp contrast to Fig. \ref%
{Delta_V8}(c,d) with SO coupling, where a strong depletion of the total
density (also for each spin component) in the dark soliton is observed. In
Fig. \ref{Delta_V8}(b), we observe two length scales for the dark soliton.
One is $K_{F}^{-1}$ defined using the local density approximation,
corresponding to the steep slope and the oscillation wavelength. The other
is the coherence length, corresponding to the smoother oscillation slope~%
\cite{Stringari2007PRA,Ho2006PRL}, which is about $%
5K_{F}^{-1},2.2K_{F}^{-1},1.2K_{F}^{-1},K_{F}^{-1}$ for $\Gamma =0.65\pi
,\pi ,1.5\pi ,2.5\pi $, respectively. These two scales are equivalent in the
BEC limit, where the oscillation structure of the soliton vanishes.

\begin{figure}[t]
\includegraphics[width=3.4in]{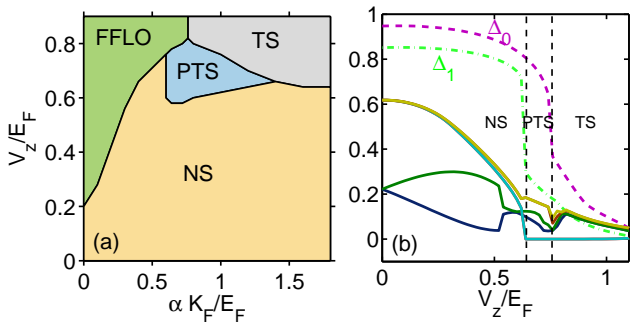}
\caption{(Color online) (a) Phase diagram in the $(\protect\alpha ,V_{z})$
plane. PTS: partial topological superfluid; TS: topological superfluid; NS:
normal superfluid. (b) Quasi-particle excitation energies (all other lines
except light green dashed-dotted and magenta dashed lines) and $\Delta _{0}$
(maximum of the order parameter in whole region), $\Delta _{1}$ (minimum of
the order parameter in region ($2x_{s}<x<9x_{s})$, where the topological
superfluids exist). $\protect\alpha K_{F}=E_{F}$ and $\Gamma =\protect\pi $.
}
\label{phase}
\end{figure}

The physical mechanism for the dark soliton in an imbalanced Fermi gas roots
in the SO coupling. 
Without SO coupling, the ground state of a 1D imbalanced Fermi gase in a
harmonic trap is predicted to possess the two-shell structure: partially
polarized core (i.e. FFLO state) surrounded by either a paired or a fully
polarized phase~\cite{HuHui2007PRA,Xiwen2013RMP}. This two-shell structure
has been partially verified in the experiment~\cite{Hulet2010Nature}. With
SO coupling, the FFLO state is dramatically suppressed as shown in Fig.~\ref%
{phase}(a) because the BCS type of zero total momentum Cooper pairing can be
formed in the same helicity band, which is energetically preferred than the
FFLO state that is formed through the pairing between atoms in two different
helicity bands~\cite{PRLTrivedi,Yong2013Arxiv} with non-zero total momentum.
A dark soliton can be created in the BCS type of phases, but not in the FFLO
phase, where the stable state generated by phase imprinting is also an
oscillating FFLO state with a sinusoidal-like form and has lower energy than
a soliton state. In this sense, SO coupling makes it possible to generate a
single dark soliton excitation with a high spin imbalance.

\begin{figure}[t]
\includegraphics[width=3.4in]{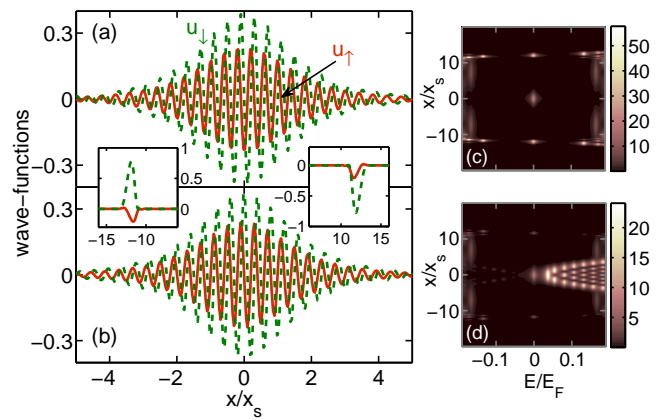}
\caption{(Color online) The left panel shows the wave-functions (solid red
line for $u_{\uparrow }$ and dashed green one for $u_{\downarrow }$) of four
MFs: two inside the dark soliton plotted in (a) and (b) and two at the edges
in the insets. (c) and (d) show the local density of states for spin $%
\downarrow $ and spin $\uparrow $ respectively. Here $V_{z}=0.768E_{F}$, $%
\protect\alpha K_{F}=E_{F}$, and $\Gamma =\protect\pi $. }
\label{MF}
\end{figure}

\emph{Two MFs inside a dark soliton}: There are two topological superfluid
(TS) phases in Fig.~\ref{phase}(a): partial TS (PTS) phase, where the
superfluid has a phase separation structure with a normal superfluid core
surrounded by a TS, and TS phase, where the whole region is topological. The
phase separation in the PTS phase originates from the harmonic trapping
geometry, where the chemical potential is replaced by a local one $\bar{\mu}%
(x)=\mu -V(x)$ using the local density approximation. For a homogeneous gas,
superfluids become topological when $V_{z}>\sqrt{{\mu }^{2}+\Delta ^{2}}$~%
\cite{Sato2009PRL,ShiLiang2011PRL} with Majarona zero modes located at the
edges. For a harmonically trapped gas, as the chemical potential decreases
from the trap center to edge, the condition $V_{z}>\sqrt{\bar{\mu}^{2}\left(
x\right) +\Delta ^{2}}$ ~\cite{XJLiu2012PRA} (See supplementary materials)
is first satisfied at the wings of the superfluids ($V_{z}<\sqrt{\mu
^{2}+\Delta ^{2}}$ at the trap center) for the PTS phase. This transition is
characterized by the appearance of two zero energy modes (cyan line and red
one hidden behind the cyan in Fig.~\ref{phase}(b)) located around the place
with $V_{z}=\sqrt{\bar{\mu}^{2}+\Delta ^{2}}$, and the sharp decrease of the
order parameter in the topological region ($\Delta _{1}$ in Fig.~\ref{phase}%
(b)). As $V_{z}$ is furthur increased with $V_{z}>\sqrt{\mu ^{2}+\Delta ^{2}}
$, the whole superfluids become topological (TS phase) as the maximum of the
order parameter ($\Delta _{0}$) drops suddenly. Without a soliton, there is
only one zero\ energy mode in TS phase. However, a soliton induces another
zero energy mode. This additional zero energy mode is different from
conventional local gapped excitations (blue and green lines in Fig.~\ref%
{phase}(b)) that are similar to Andreev bound states in a vortex~\cite%
{Ho2006PRL,Stringari2007PRA}. It appears only when the local superfluid,
where the soliton is located, becomes topological.

To confirm that the additional zero energy state is MFs inside the soliton,
we consider a linear combination of the Bogoliubov quasi-particle operators $%
\gamma _{0^{n}}$ for states with $E_{n}\sim 0$ (here $E_{2}>E_{1}$
correspond to cyan and red lines respectively) to obtain spatially localized
states: $\gamma _{L}=(\gamma _{0^{+2}}+\gamma _{0^{+1}}+\gamma
_{0^{-2}}+\gamma _{0^{-1}})/2$, $\gamma _{R}=(\gamma _{0^{+2}}-\gamma
_{0^{+1}}+\gamma _{0^{-2}}-\gamma _{0^{-1}})/2$, $\gamma _{S1}=(\gamma
_{0^{+2}}-\gamma _{0^{-2}})/2i$, $\gamma _{S2}=(\gamma _{0^{+1}}-\gamma
_{0^{-1}})/2i$. Due to the particle-hole symmetry $\gamma _{0^{n}}=\gamma
_{0^{-n}}^{\dagger }$, we obtain $\gamma _{L}^{\dagger }=\gamma _{L}$, $%
\gamma _{R}^{\dagger }=\gamma _{R}$, and $\gamma _{S\sigma }^{\dagger
}=\gamma _{S\sigma }$ with $\sigma =1,2$, indicating that $\gamma _{L}$, $%
\gamma _{R}$ and $\gamma _{S\sigma }$ are self Hermitian Majorana operators.
In Fig.~\ref{MF}(a,b), we plot the wave-functions of MFs, showing that two
MFs at the left and right edges (two insets) and two MFs inside the soliton
((a) and (b)). The wave-functions of MFs inside the soliton behave like $%
\sim \cos (\pi K_{F}x/2)\exp (-x^{2}/\xi _{0}^{2})$ and $\sim \sin (\pi
K_{F}x/2)\exp (-x^{2}/\xi _{0}^{2})$, similar to Andreev bound states~\cite%
{Stringari2007PRA} but with $\xi _{0}$ larger than $\xi $. They are very
different from widely known MFs in vortices or nanowire ends that interact
due to the wave-function overlap, leading to energy splitting that destroys
the zero energy states. The vanishing interaction between MFs inside the
dark soliton is due to the intrinsic property of a dark soliton: a sharp
phase change. We can understand this through Kitaev's toy model~\cite%
{Kitaev2001}, and the interaction is proportional to $-i\frac{t}{2}\cos
(\delta \phi /2)\gamma _{S1}\gamma _{S2}$ if written in the Majorana fermion
representation with the phase difference $\delta \phi $ between two sides of
the soliton. For a dark soliton, $\delta \phi =\pi $ and there is no
interaction. Such coexistence of two MFs inside a soliton makes it possible
to drag MFs to a new position given that a dark soliton can be manipulated
by means of optical lattices~\cite{Theocharis2005PRE}.

\begin{figure}[t]
\includegraphics[width=2.8in]{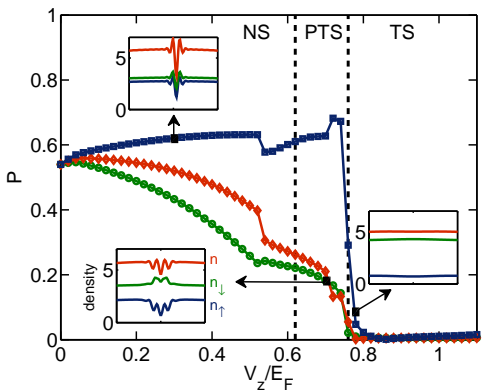}
\caption{(Color online) Density contrast $P_{\protect\sigma }$ as a function
of $V_{z}$ for the spin $\downarrow $ (circle green line), spin $\uparrow $
(square blue line), and total (diamond red line) densities in the soliton
region $-1.5x_{s}<x<1.5x_{s}$. The insets show the density profiles ($n_{%
\protect\sigma }$ for spin $\protect\sigma $ and $n=n_{\uparrow
}+n_{\downarrow }$) of the soliton in each region with the unit $1/x_{s}$,
corresponding to $V_{z}=0.3E_{F},0.7E_{F},0.768E_{F}$ respectively. Here $%
\protect\alpha K_{F}=E_{F}$ and $\Gamma =\protect\pi $.}
\label{density_constrast}
\end{figure}

\emph{Evolution of soliton structure}: The soliton structure can be
characterized by the density contrast $P_{\sigma }=(n_{\text{max}}^{\sigma
}-n_{\text{min}}^{\sigma })/n_{\text{max}}^{\sigma }$ with the maximum $n_{%
\text{max}}^{\sigma }$ and minimum $n_{\text{min}}^{\sigma }$ of the density
for spin $\sigma $ ($P_{t}$ for the total density) in the soliton region. In
Fig.~\ref{density_constrast}, we plot $P_{\sigma }$ as a function of $V_{z}$%
. We see both $P_{\downarrow }$ and $P_{t}$ decrease, while $P_{\uparrow }$
is almost a constant with increasing $V_{z}$ before the appearance of MFs
inside the soliton in the TS region. However, the soliton structure almost
vanishes in the TS region with two MFs accommodating the soliton. The sudden
decrease of the density contrast at the topological transition point is
mainly due to the sharp decrease of the pairing order parameter, as shown by
$\Delta _{0}$ in Fig.~\ref{phase}(b), leading to the sudden decrease of the
number of atoms participating in the pairing around the soliton. The sudden
disappearance of the density contrast across the dark soliton might provide
an experimental signature for the appearance of MFs inside the dark soliton.
From the insets, we see that the soliton density $n_{\downarrow }$ has a
convex structure for $V_{z}>0.53E_{F}$, where the density inside the soliton
is larger than its surrounding. The transition to this convex structure
leads to the kinks around $V_{z}=0.53E_{F}$ observed in Fig.~\ref{phase}(b)
and Fig.~\ref{density_constrast}. This convex structure of soliton density
is caused by the local quasi-particle excitations. In fact, without taking
into account of such quasi-particle contributions to the density, $%
P_{\downarrow }$ is almost zero.

\emph{Experimental signature of topological solitons and MFs}: In the right
panel of Fig.~\ref{MF}, we present the local density of states~\cite%
{HuHui2007PRA} of the Fermi gas, $\rho _{\sigma }(x,E)=\sum_{|E_{n}|<E_{c}}%
\left[ |u_{n\sigma }|^{2}\delta (E-E_{n})+|v_{n\sigma }|^{2}\delta (E+E_{n})%
\right] /2$ with $\sigma =\uparrow ,\downarrow $, which reflect the zero
energy excitations at the places where MFs are locally accommodated.
Clearly, in the TS region, the zero energy MF states appear at $x=0$, where
the dark soliton locates. In experiments, the LDOS could be measured using
spatially resolved radio-frequency (rf) spectroscopy~\cite{Shin2007PRL} with
the space resolution about $1.4\mu m$ and spectral resolution about $0.5kHz$
that are smaller than the space resolution $5.4\mu m$ and energy resolution $%
3kHz$ shown in Fig.~\ref{MF}(c, d)~\cite{Hulet2010Nature}. The LDOS in a
dark soliton in the trap center provides a stronger and stabler signal than
that at the edges with small atom densities around MFs~\cite{XJLiu2012PRA}.

The 1D SO coupling and Zeeman fields considered here have already been
achieved for $^{40}$K and $^{6}$Li fermionic atoms by coupling two hyperfine
ground states using two Raman laser beams in experiments~\cite%
{Jing2012PRL,Zwierlen2012PRL,Spilman2013PRL}. In experiment, the SO coupling
and Zeeman field strength can be tuned through varying the laser intensity
or the setup of the laser beams. The SO coupling can be as large as $\alpha
K_{F}\sim E_{F}$ and a Zeeman field can be readily tuned to $V_{z}\sim E_{F}$%
. The realization of 1D Fermi gases and the dark soliton can be similar as
that in recent experiments \cite{Martin2013Nature}, where an elongated 1D
Fermi gas is confined in a harmonic trap with cylindrical symmetry (radial
trapping frequency much larger than axial one) using a combination of weak
magnetic trap (axial) and tightly focused optical trap (radial). Dark
solitons can be experimentally created via phase imprinting~\cite%
{Burger1999PRL,Denschlag2000Science,Martin2013Nature}, where a half of the
cloud is shortly interacted with a laser beam to acquire the phase
difference.

\emph{Discussion}: The mean-field BdG theory used here may give a
qualitative description of the 1D physics. In particular, for 1D Fermi gases
with weak and moderate interactions, the energy and chemical potential
obtained from mean-field theory and exact Bethe ansatz were compared \cite%
{HuHui2007PRA} and only small discrepancy was found. Moreover, the
fluctuations could be suppressed in an experimentally trapped Fermi gas,
where the density of states is a constant, similar to homogenous 2D systems
\cite{HuHui2007PRA}. For a SO coupled Fermi gas, recent comparison between
the mean-field phase diagram and the exact 1D DMRG simulation shows the
qualitative correctness of the mean-field approximation~\cite{GangChenArxiv}%
. Furthermore, quasi-1D Fermi gases can be engineered via an externally
imposed strong optical lattice in experiments~\cite{Hulet2010Nature} and the
weak tunneling ~\cite{Parish2007PRL,Kuei2012PRA,Qu2014MChain} along
transverse directions can be tuned to suppress the fluctuations. For
solitons in highly elongated DFGs, the experimentally observed long period
of the soliton oscillation was found to be in good agreement with the
hydrodynamic theory \cite{Martin2013Nature}, which is an approximation of
the mean-field BdG approach. Our work provides the first step approach to
understand the fundamental soliton physics in this system, and more
quantitative results need further investigation.

To conclude, we showed that a single dark soliton excitation can exist in an
imbalanced DFG with SO coupling, in sharp contrast to the FFLO state without
SO coupling. With a substantial spin imbalance, we found that two MFs can
coexist inside one single dark soliton without interactions, which provides
a new avenue for experimentally observing and manipulating MFs by
controlling solitons as well as creating Majorana trains by engineering
soliton trains~\cite{Peter2011PRL}.

\textit{Acknowledgement:} We would like to thank X. Liu, R. Chu, and M. J.
H. Ku for helpful discussions. Y. Xu, L. Mao, and C. Zhang are supported by
ARO(W911NF-12-1-0334), AFOSR (FA9550-13-1-0045), and NSF-PHY (1249293). L.
Mao is also supported by the NSF of China (11344009). B. Wu is supported by
the NBRP of China (2013CB921903,2012CB921300) and the NSF of China
(11274024,11334001). We also thank Texas Advanced Computing Center, where
our program is performed.

\begin{widetext}
\section{Supplementary Materials}
In the main text, we show that there are four phases: NS, FFLO, PTS, and TS,
instead of three in a homogeneous space. This extral PTS phase arises from
the existence of a harmonic trap, where the chemical potential is replaced
by a local one $\bar{\mu}=\mu -V(x)$ with the local density approximation.
It is well known that a homogeneous spin-orbit coupled superfluid becomes
topological when $V_{z}>\sqrt{\mu ^{2}+\Delta ^{2}}$. In a harmonic trap,
this relation can be approximately written as $V_{z}>\sqrt{\bar{\mu}%
^{2}+\Delta (x)^{2}}$. Based on this relation, a superfluid can have a phase
separation structure (i.e. PTS): a topological superfluid wing ($V_{z}>\sqrt{%
\bar{\mu}^{2}+\Delta (x)^{2}}$ because of the smaller effective chemical
potential $\bar{\mu}$) surrounding a normal superfluid core ($V_{z}<\sqrt{%
\mu ^{2}+\Delta (x)^{2}}$). This phase exists when $|\Delta |<V_{z}<\sqrt{%
\mu ^{2}+\Delta (0)^{2}}$. In Fig.~\ref{Sub}, we present the order parameter
profile $|\Delta (x)|$ and $\Delta V=V_{z}-\sqrt{\bar{\mu}^{2}+\Delta ^{2}}$
in a large region for three phases: (a) NS; (b) PTS; (c) TS. In the NS
phase, $V_{z}$ is so small that $V_{z}<\sqrt{\bar{\mu}^{2}+\Delta (x)^{2}}$
in the whole space and there are no topological superfluids. As $V_{z}$ is
increased that $V_{z}>\sqrt{\bar{\mu}^{2}+\Delta (x)^{2}}$ in the two shells
as shown in Fig.~\ref{Sub} (b) so that superfluids become partially
topological associated with two zero energy states. This transition can also
be characterized by a sharp decrease of the minimum of the order parameter
(i.e. $\Delta _{1}$) at the topological superfluids region ($2x_{s}<x<9x_{s}$%
) as illustrated in Fig. 2(b) in the main text. As $V_{z}$ is furthur
increased, $V_{z}>\sqrt{\mu ^{2}+\Delta (x)^{2}}$ and the whole superfluids
become topological. In this phase, a soliton in the trap center hosts two
Majorana fermions. Across the transition, the maximum of the order parameter
($\Delta _{0}$) drops suddenly as shown in Fig. 2(b) in the main text.

\setcounter{figure}{0} \renewcommand{\thefigure}{S\arabic{figure}}
\begin{figure}[t]
\includegraphics[width=3.4in]{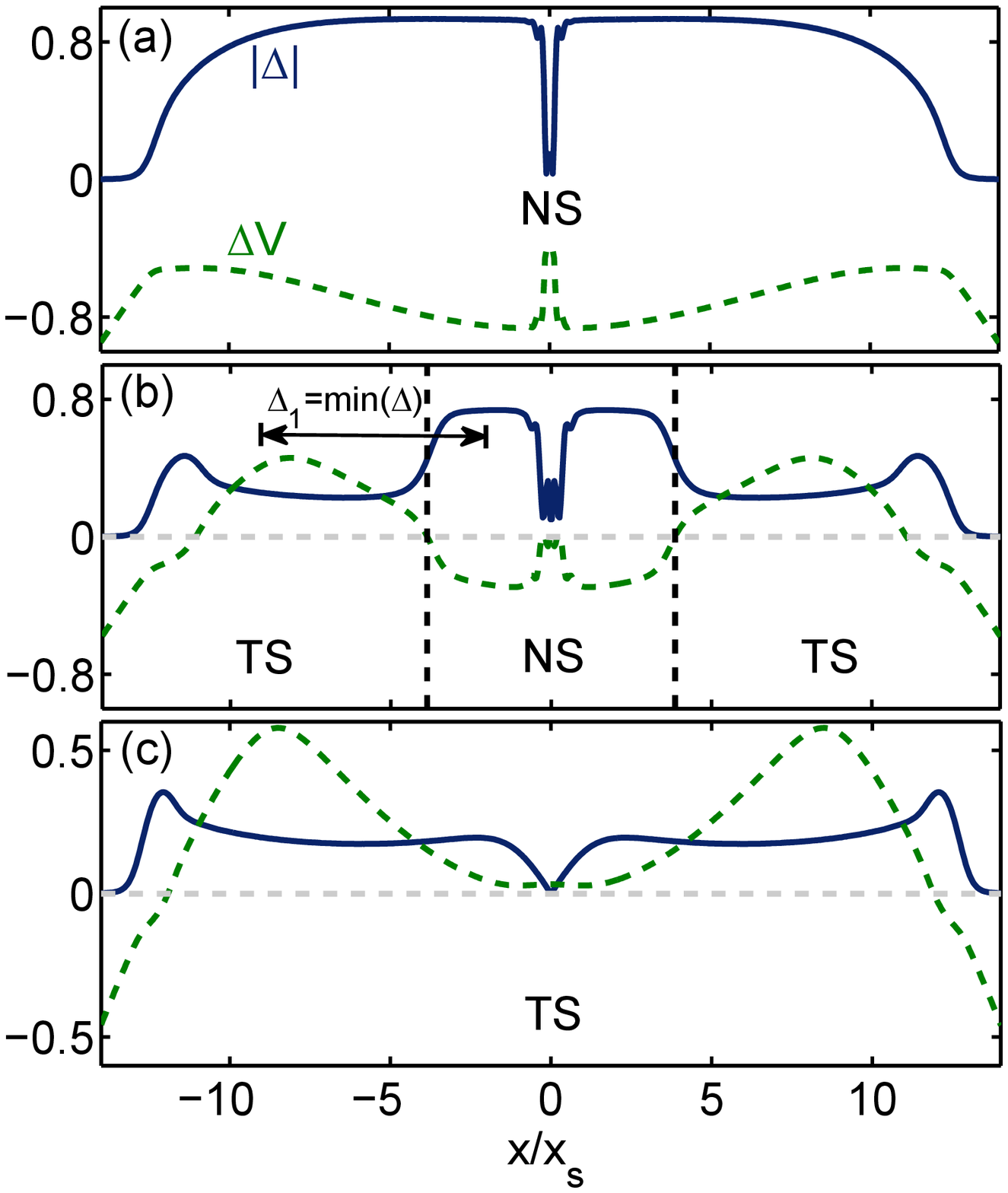} .
\caption{(Color online) Plot of the order parameter profile $|\Delta (x)|$
and $\Delta V=V_{z}-\protect\sqrt{\bar{\protect\mu}^{2}+\Delta (x)^{2}}$ in
(a) for the NS phase with $V_{z}=0.3E_{F}$, in (b) for the PTS phase with $%
V_{z}=0.7E_{F}$, and in (c) for the TS phase with $V_{z}=0.768E_{F}$. Here $%
\protect\alpha K_{F}=E_{F}$ and $\Gamma =\protect\pi $. $\Delta _{1}$ is the
minimum of the order parameter in region $2x_{s}<x<9x_{s}$. The depletion at
the center originates from the existence of a soliton.}
\label{Sub}
\end{figure}

\end{widetext}
\end{document}